\documentclass[12pt]{iopart}
\usepackage{iopams}
\usepackage{graphicx}
\usepackage{color}
\usepackage{amsfonts}
\input amssym.def 
\input amssym

\begin{document}

FQXi Essay Contest - It from Bit or Bit from It? June 2013

\title[]{ It from qubit: how to draw quantum contextuality}

\author{ Michel Planat}

\vspace*{.1cm}
\address{$^1$Institut FEMTO-ST, CNRS, 32 Avenue de
l'Observatoire, 25044 Besan\c con, France.}
\ead{michel.planat@femto-st.fr}

\begin{abstract}

Wheeler's {\it observer-participancy} and the related {\it it from bit} credo refer to quantum non-locality and contextuality. The mystery of these concepts slightly starts unveiling if one encodes the (in)compatibilities between qubit observables in the relevant finite geometries. The main objective of this treatise is to outline another conceptual step forward by employing Grothendieck's {\it dessins d'enfants} to reveal the topological and (non)algebraic machinery underlying the measurement acts and their information content. 

\end{abstract}


\section{Introduction}
\noindent

I can no better summarize the topic of this essay than by borrowing the opinions of three giants of physics, namely of J. A. Wheeler on the \lq it from bit', of J. S. Bell on \lq nonlocality' and of N. D. Mermin on \lq contextuality' (I could also have referred to A. Peres).

Wheeler \cite[(a)]{Wheeler}: {\it We have clues, clues most of all in the writings of Bohr, but not answer ... Are billions upon billions of acts of observer-participancy the foundation of everything? We are about as fas as we can today from knowing enough about the deeper machinery of the universe to answer this question. Increasing knowledge about detail has brought an increasing ignorance about the plan.} 

Wheeler \cite[(b)]{Wheeler}: {\it It from bit. Otherwise put, every it - every particle, every field of force, even the space time continuum itself - derives its function, its meaning, its very existence entirely - even if in some contexts indirectly - from the apparatus-elicited answers to yes or no questions, binary choices, bits}. 

Bell \cite{Bell1964}: {\it In a theory in which parameters are added to quantum mechanics to determine the results of individual measurements, without changing the statistical predictions, there must be a mechanism whereby the setting of one measuring device can influence the reading of another instrument, however remote. Moreover, the signal involved must propagate instantaneously, so that a theory could not be Lorentz invariant.}

Mermin \cite[(a)]{Mermin1993}: {\it It is also appealing to see the failure of the EPR reality criterion emerge quite directly from the one crucial difference between the elements of reality (which, being ordinary numbers, necessarily commute) and the precisely corresponding quantum mechnical observables (which sometimes anticommute)}.

On the mathematical side, my arguments will rely on another eminent figure of science, viz.

Grothendieck \cite[(a), Vol. 1]{Schneps1}: {\it In the form in which Belyi states it, his result essentially says that every algebraic curve defined over a number field can be obtained as a covering of the projective line ramified only over the points $0$, $1$ and $\infty$. The result seems to have remained more or less unobserved. Yet it appears to me to have considerable importance. To me, its essential message is that there is a profound identity between the combinatorics of finite maps on the one hand, and the geometry of algebraic curves defined over number fields on the other. This deep result, together with the algebraic interpretation of maps, opens the door into a new, unexplored world - within reach of all, who pass by without seeing it}.

Why do I refer to A. Grothendieck? Below is a sketch of an ongoing work \cite[a]{Planat2013}. My first point is that Wheeler's observer-participancy is contextual: the {\it it} does not preexist to the measurement set-up and -- the {\it it from bit} extraction being even more intriguing -- the measured value depends on all mutually compatible measurements. Second, a compatibility (i.\,e., commutativity) diagram of observables itself has a kind of engine that drives it. The hidden engine is, in my opionion, nothing but Grothendieck's {\it dessin d'enfant} (a child's drawing). The relevant `quantum' graphs  that we will encounter on our journey are mainly commuting/anticommuting graphs discovered by N. D. Mermin. We will ask ourselves the question whether, and how, a {\it dessin d'enfant} (an algebraic curve in the sense of Grothendieck) can be associated with the contextual set-up behind such a graph. Interestingly enough, also confirming the {\it it from bit} claim, any relevant  `non-algebraic' graph will be seen to have more Shannon capacity/information than expected in a perfect graph.  This sounds in resonance with Shor's quantum algorithm that allows factoring of integers in polynomial time instead of exponential one (more information in less time!).

It will suffice to play with the so-called Pauli groups of operators/observables for two or three parties, i.\,e. two- or three-qubit systems: Alice, Bob and/or Charlie set-ups. Most importantly, experiments with compatible (mutually commuting) operators make sense, otherwise the results are expected to be fully independent \cite[(b)]{Mermin1993}. 
Some years ago it was realized that to mathematically grasp the essence of these (in)compatibilies finite geometries have to be called on \cite[(b)]{Planat2013}. In what follows, it will be demonstrated that to get further insights in this respect,  the concept of a {\it  dessin d'enfant} -- a bipartite graph embedded on an oriented surface -- must enter the game. Any {\it dessin} can be given the structure of a Riemann surface, and the Riemann surfaces arising this way are those defined over the field of algebraic numbers (Belyi's theorem). This subject is briefly described in Sec. \ref{Belyi}.

If one considers the fifteen two-qubit operators (the identity matrix is discarded), here the underlying geometry -- the smallest non-trivial generalized quadrangle -- tells us a lot about {\it higgledy-piggledy} collection of potential two-qubit experiments \cite[(b)]{Planat2013}. But, as emphasized by Mermin \cite[(b)]{Mermin1993}, to unify all relevant concepts pertinent to non-locality and contextuality one needs three parties. Here, a fundamental building block is a heptad of mutually commuting operators forming the smallest projective plane, the Fano plane, having $7$ points and dually $7$ lines, with three points per line and three lines through a point.
In Sec. \ref{Fano}, the seven points of the Fano plane are put in a bijective correspondence with the edges of a tree-like {\it dessin d'enfant}, ensuring its algebraicity (in Grothendieck's sense). The symmetry group permuting the edges of the {\it dessin} and stabilizing the lines of the Fano plane is the well known simple group of cardinality $168$.

Then, in Sec. \ref{Bell}, I invoke a square graph with eight vertices, with mutually commuting or anticommuting operators, for a proof of Bell's theorem about non-locality. Here, the graph is itself the {\it dessin}, and is thus algebraic. Sec. \ref{Mermin} starts a non-trivial story about contextuality and the so-called Kochen-Specker theorem as told by N. D. Mermin. We only need a two-player (two-qubit) set-up. The graph is a `magic' square (a three-by-three grid) embodying a contradiction between the algebra of operators and eigenvalues \cite[(b)]{Mermin1993}. We shall uncover in its shadow an interesting algebraic curve/{\it dessin d'enfant} -- displayed here for the first time. The symmetry group permuting the nine edges of the {\it dessin} and stabilizing the lines of the magic square has order $72$. Finally, in Sec. \ref{Merminagain}, it comes -- as expected -- both contextual and `magic' array of three-qubit observables, the so-called Mermin's  pentagram. Apart from being non-algebraic, in Grothendieck's sense, it also occurs in another disguise: the Petersen graph, being thus embeddable as a polyhedron in the real projective plane. Such a pentagram is one of a totality of $12096$ guys fitting the structure of the smallest exceptional Lie geometry -- the split Cayley hexagon of order two \cite[(c)]{Planat2013}.

\section{What is a dessin d'enfant?}
\label{Belyi}

{\it ... here I was brought back, via objects so simple that a child learns them while playing, to the beginnings and origins of algebraic geometry, familiar to Riemann and his followers! \cite[(a), vol. 1]{Schneps1}.}

Details can be found in \cite{Schneps1,Lando2004}. See also \cite{Koch2010} for a link to Feynman diagrams.

* Step 1 (easy): A {\it dessin d'enfant} (child's drawing)   is a map drawn on a surface (a smooth compact orientable variety of dimension two) such that vertices are points, edges are arcs connecting the vertices, and the complement of the graph is the union of faces (each one homeomorphic to the open disk of $\mathbb{R}^2$). The graph may feature multiple edges and/or loops, but has to be connected. Taking $S$, $A$ and $F$ for the number of vertices, edges and faces, respectively, the genus $g$ of the map is given by Euler's formula $S-A+F=2-2g$. 

* Step 2 (still easy): Convert the graph of Step 1 into a bipartite graph (called also a hyper-map) by regarding its vertices as black and placing a white vertex in each of its edges. The set of half-edges so defined is encoded by a permutation group $P=\left\langle \alpha, \beta \right\rangle$; here, the permutation $\alpha$ (resp. $\beta$) rotates the half-edges around each black (resp. white) vertex in accord with the cyclic ordering in that vertex. The cycle structure for the faces follows from the permutation $\gamma$ satisfying $\alpha\beta\gamma=1$. The Euler characteristic now reads $2-2g=B+W+F-n$, where $B$, $W$ and $n$ stands, respectively, for the number of black vertices, the number of white vertices and the number of half-edges.  

It is also known that maps on connected oriented surfaces are parametrized by the conjugacy classes of subgroups of the triangle group, also called cartographic group by A. Grothendieck,

\begin{equation}
C_2^+=\left\langle  \rho_0,\rho_1,\rho_2|\rho_1^2=\rho_0\rho_1\rho_2=1 \right\rangle.
\label{cartographic}
\end{equation}
The possible existence of a {\it dessin d'enfant} of prescribed properties (permutation group, cycle structure) can thus be checked from a systematic enumeration of conjugacy classes of $C_2^+$.

* Step 3 (difficult): The Belyi theorem states that to a combinatorial map (defined in Step 2) there corresponds a Riemann surface $X$ defined over the field $\bar{\mathbb{Q}}$ of algebraic numbers. This happens if, and only if, there exists a covering $f: X\rightarrow \bar{\mathbb{C}}$ unramified outside $\{0,1,\infty\}$.
The covering (an algebraic curve) $f$ associated with a {\it dessin d'enfant} is, in general, very difficult to find explicitly, except for simple cases devoid of loops. Klein, as early as 1884, was the first to get them for the graphs representing Platonic solids  \cite{Klein}.

\subsection*{The Fano plane and its curve} 
\label{Fano}
\noindent

A finite projective plane is a point-line incidence geometry such that (i) any two lines meet in a unique point, (ii) any two distinct points are on a unique line, and (iii) there are  at least four points, not three of them collinear. The simplest case is the projective plane of order two -- the Fano plane. Projective planes are conjectured to exist only if their order is a power of a prime number.

As already mentioned, a three-qubit maximal commuting set may be seen an heptad of points/lines satisfying the axioms of a Fano plane \cite{Levay2008}. 
The seven points are mapped to the seven three-qubit operators and each line features a triple of mutually commuting operators whose product is $\pm$ the identity matrix.
We are in the search of a {\it dessin d'enfant}  $\mathcal{D}_\mathcal{F}$ whose edges can be put in a bijective correspondence with the points of the Fano plane $\mathcal{F}$ in such a way that the permutation group $P$ of $\mathcal{D}_\mathcal{F}$ acts transitively on the lines of $\mathcal{F}$ and stabilize them. One knows that $P=PSL(2,7)$, the simple group of order $168$. Using the software Magma, we are able to compute the $131$ subgroups of index $7$ of the cartographic group $C_2^+$, extract the $10$ of them whose group of cosets is isomorphic to $P$ and also have the right action on $\mathcal{F}$. One choice is depicted in Fig.\,1, as reproduced from \cite[(a), vol. 2,  p. 17 and p. 50]{Schneps1};
the corresponding Belyi map is the polynomial
$$Z=z^4(z-1)^2(z-a)~~\mbox{with}~a=(-1-i\sqrt{7})/4.$$

\begin{figure}
\centering 
\includegraphics[width=6cm]{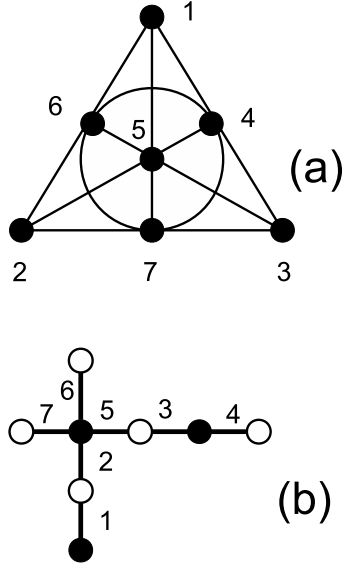}
\caption{ (a) The Fano plane $\mathcal{F}$ and (b) a corresponding {\it dessin d'enfant} $\mathcal{D}_{\mathcal{F}}$. The permutation group characterising $\mathcal{D}_{\mathcal{F}}$ is $P=\left \langle \alpha,\beta \right \rangle$, with $\alpha=(1)(2,7,6,5)(3,4)$ and $\beta=(1,2)(3,5)(4)(6)(7)$. One has $B=3$ black vertices, $W=5$ white vertices, $n=7$ half-edges, $F=1$ face and genus $g=0$.}
\label{fig1}
\end{figure}

\section{Dessins d'enfants, non-locality and contextuality}

\subsection{A {\it dessin d'enfant} for Bell's theorem} 
\label{Bell}

Let us recall now that for dichotomic observables $\sigma_i^{2}=\pm 1$, $i = 1,2,3,4$, when giving the pair $(\sigma_1,\sigma_3)$ to Bob and the pair $(\sigma_2,\sigma_4)$ to Alice, the Bell-CHSH approach consists of defining the number 
$$C=\sigma_2(\sigma_1+\sigma_3)+\sigma_4(\sigma_3-\sigma_1)=\pm 2$$
and observing the Bell-CHSH inequality \cite[p. 164]{Peres}
$$|\left\langle \sigma_1\sigma_2 \right\rangle+\left\langle \sigma_2\sigma_3 \right\rangle+\left\langle \sigma_3\sigma_4 \right\rangle-\left\langle \sigma_4\sigma_1 \right\rangle|\le 2,$$
where $\left\langle  \right\rangle$ here means that we are taking averages over many experiments.
Bell's theorem simply means finding a violation of the afore-mentioned inequality with quantum observables and dichotomic eigenvalues. A simple choice is the quadruple
\begin{equation}
(\sigma_1=IX,~\sigma_2=XI,~\sigma_3=IZ,~\sigma_4=ZI),
\label{quadruple}
\end{equation}
where $X$, $Y$ and $Z$ are the ordinary Pauli spin matrices and one uses the short-hand notation for the tensor product, e.\,g. $IX \equiv I \otimes X$. Thus $\sigma_i^2=1$, and one finds that 
$$C^2=4*I +[\sigma_1,\sigma_3][\sigma_2,\sigma_4]=4 \left(\begin{array}{cccc} 1 &. & . &1 \\ . &1 & \bar{1} &.  \\ . &\bar{1} & 1&. \\ 1 &. & . &1   \end{array}\right)$$ 
has eigenvalues $0$ and $8$, both of multiplicity $2$. Taking the norm of the bounded linear operator $A$ as $||A||=sup (||A \psi||/||\psi||),~\psi \in \mathcal{H}$ (the relevant Hilbert space), one gets the maximal violation of the Bell-CHSH inequality \cite[p. 174]{Peres}
$$||C||=2\sqrt{2}.$$ 

\begin{figure}
\centering 
\includegraphics[width=6cm]{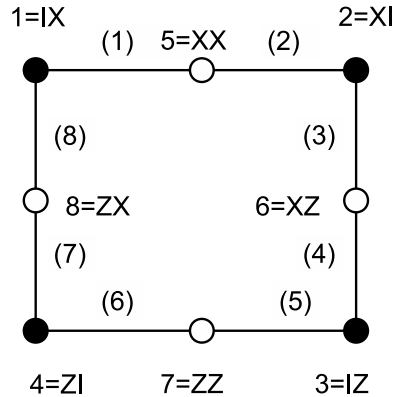}
\caption{The commutation/anti-commutation diagram for Bell's theorem is, at the same time, a {\it dessin d'enfant} with permutation group $P=\left\langle \alpha,\beta \right\rangle =D_4$, where $\alpha=(1,8)(2,3)(4,5)(6,7)$, $\beta=(1,2)(3,4)(5,6)(7,8)$ and the bracketed half-edge labelling.}
\label{fig2}
\end{figure}
\noindent
A straightforward computer check shows that there are $90$ such distinct proofs of Bell's theorem with two-qubit operators and as many as $30240$ ones with three-qubit ones, all of them yielding a maximal violation of the Bell-CHSH inequality. These numbers are intimately connected with the structure of the corresponding contextual spaces. 

Let us represent the commutation relations between the elements of (\ref{quadruple}) as illustrated in Fig.\,2, where the black vertices are labelled by $i\equiv \sigma_i$, and a white vertex represents the operator which is the product of the two operators at the endpoints of the corresponding edge (e.\,g.  $5 = \sigma_1 \sigma_2= IX.XI =XX$); it is worth noting that three operators placed along the same straight-line segment mutually commute, as do two `white'  operators situated opposite each other. This is a remarkable instance where the commutation/anti-commutation diagram is bipartite and, as it stands, it also represents a {\it dessin d'enfant} (with permutation group equal to the dihedral group $D_4$ on $8$ elements). 
The algebraic curve (Belyi map) associated to it is well known (it was already derived by Klein \cite[p. 106]{Klein})
$$Z=\frac{(z^4-1)^2}{-4z^4}.$$

\begin{figure}
\centering 
\includegraphics[width=7cm]{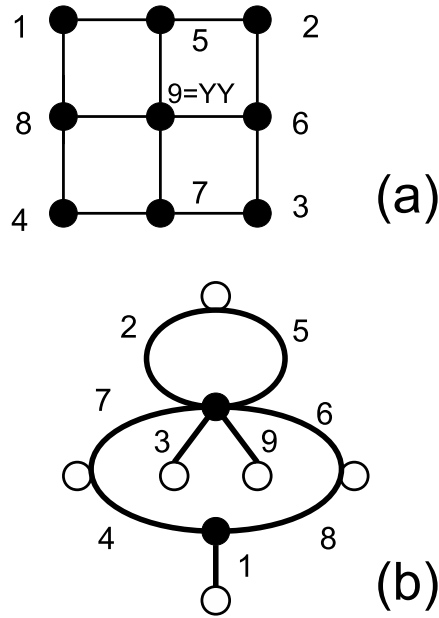}
\caption{(a) The Mermin square $\mathcal{M}$ and (b) its associated {\it dessin d'enfant} $\mathcal{D}_\mathcal{M}$ with permutation group $P=\left\langle \alpha,\beta\right\rangle$, where $\alpha=(3,9,6,5,2,7)(1,8,4))$, $\beta=(1)(3)(9)(2,5)(4,7)(6,8)$. One has $B=2$ black vertices, $W=6$ white vertices, $N=9$ half-edges, $F=3$ faces and genus $g=0$. }
\label{fig3}
\end{figure}

\subsection{Mermin's square}
\label{Mermin}

Let's have now a bit more careful look at Fig.\,2. We observe that the product of the two observables associated with a pair of the opposite white vertices is the same, $YY$. By supplying this `missing' vertex and the two lines passing through it, we get a $3 \times 3$ grid (as illustrated in Fig.\,3a). This grid is a remarkable one:  all triples of observables located in a row or a column have their product equal to   $+II$ {\it except for} the middle column, where $XX.YY.ZZ=-II$. Mermin was the first to observe that this is a Kochen-Specker (parity) type contradiction since the product of all triples  yields the matrix $-II$, while the product of corresponding eigenvalues is $+1$ (since each of the latter occurs twice, once in a row and once in a colmumn) \cite[(b)]{Mermin1993}. Such a Mermin `magic' square may be used to provide many contextuality proofs from the vectors shared by the maximal bases corresponding to a row/column of the diagram. The simplest, so-called $(18,9)$ one ($18$ vectors and $9$ bases) has, remarkably, the orthogonality diagram which is itself a Mermin square ($9$ vertices for the bases and $18$ edges for the vectors) \cite[(c), eq. (6)]{Planat2013}.

Now, we would like to represent our Mermin square, $\mathcal{M}$, in a way analogous to what we did for the Fano plane in Sec. \ref{Fano}, that is we would like to draw a {\it dessin d'enfant} $\mathcal{D}_\mathcal{M}$ whose $9$ half-edges are in a bijective correspondence with the vertices of $\mathcal{M}$ and whose permutation group $P$ acts transitively on the rows/columns of $\mathcal{M}$ by stabilizing them. The symmetry group of $\mathcal{M}$ is isomorphic to $\mathbb{Z}_3^2 \rtimes \mathbb{Z}_2^3$, a group of order $72$. Using again the software Magma, we serached for all $1551$ subgroups of index $9$ in the cartographic group $C_2^+$, defined in (\ref{cartographic}), extracted the two subgroups isomorphic to $P$ and selected the one having the right action on $\mathcal{M}$; the corresponding {\it dessin d'enfant} $\mathcal{D}_{\mathcal{M}}$ is shown in Fig. 3b. To find the corresponding Belyi map seems to be a challenging math problem.

\subsection{Mermin's pentagram} 
\label{Merminagain}
\noindent

Poincar\'e \cite[p. 342]{Flanagan}; {\it Peceptual space is only an image of geometric space, an image altered in shape by a sort of perspective.}

Weyl \cite[p. 343]{Flanagan}: {\it In this sense the projective plane and the color continuum are isomorphic with one another.}

Color experience through our eyes to our mind relies on the real projective plane $\mathbb{R}\mathbb{P}^2$ \cite{Flanagan}. Three-qubit contextuality also relies on $\mathbb{R}\mathbb{P}^2$ thanks to a Mermin `magic' pentagram, that for reasons explained below in (i) we denote $\bar{\mathcal{P}}$ (by abuse of language because we are at first more interested to see the pentagram as a geometrical configuration than as a graph). 
One such a pentagram is displayed in Fig. 4a. It consists of a set of five lines, each hosting four mutually commuting operators and any two sharing a single operator. The product of operators on each of the lines is $-III$, where $I$ is the $2 \times 2$ identity matrix. It is impossible to assign the dichotomic truth values $\pm 1$ to eigenvalues while keeping the multiplicativve properties of operators so that the Mermin pentagram is, like its two-qubit sibling, `magic', and so contextual \cite[a]{Mermin1993},\cite[(b) and c)]{Planat2013}.

Let us enumerate a few remarkable facts about a pentagram. 

(i) The graph $\bar{\mathcal{P}}$ of a pentagram is the complement of that of the celebrated Petersen graph $\mathcal{P}$. One noticeable property of $\mathcal{P}$ is to be the smallest bridgeless cubic graph with no three-edge-coloring. The Petersen graph is thus not planar, but it can be embedded without crossings on $\mathbb{R}\mathbb{P}^2$ (one of the simplest non-orientable surfaces), as illustrated in Fig 4b.

(ii) There exist altogether $12096$ three-qubit Mermin pentagrams, this number being identical to that of automorphisms of the smallest split Cayley hexagon $G_2(2)$ -- a remarkable configuration of $63$ points and $63$ lines, whose structure is fully encoded in that of the Fano plane \cite[(b)]{Planat2013}.

\begin{figure}
\centering 
\includegraphics[width=7cm]{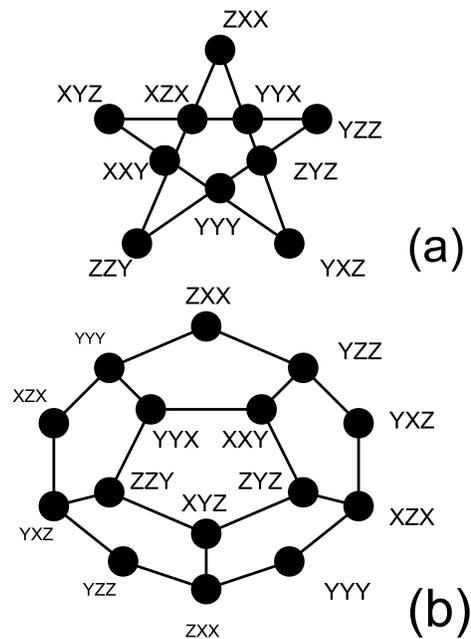}
\caption{(a) A Mermin pentagram $\bar{\mathcal{P}}$ and (b) the embedding of the associated Petersen graph $\mathcal{P}$ on the real projective plane as a hemi-dodecahedron.}
\label{fig4}
\end{figure}

(ii) Now comes an item close to the {{\it it from bit} perspective, if one employs the so-called Shannon capacity of $\bar{\mathcal{P}}$. The Shannon capacity $\Theta(G)$ of a graph $G$ is the maximum number of $k$-letter messages than can be sent through a channel without a risk of confusion. One knows that $\Theta(G)$ has for its lower bound the size $\alpha(G)$ of a maximum independent set and for its upper bound the  Lovasz number $\theta(G)$. For the complement graph $\bar{G}$, the Lovasz number $\theta(\bar{G})$ is found to lie between the clique number $\omega(G)$  and the chromatic number $\kappa(G)$. For the Petersen graph, this leads to $2\le \Theta(P) \le 4$ and for the pentagram graph $2\le \theta(\bar{P}) \le 3$.
A direct calculation yields $\Theta(\bar{P}) \ge \sqrt{5}$ \cite[(c)]{Planat2013}. (Note hat the pentagon graph attains the tight bound $\sqrt{5}$.)

(iv) Does it exist a {\it dessin d'enfant} for the pentagram? In view of the relationship of $\bar{P}$ to  the non-orientable $\mathbb{R}\mathbb{P}^2$, this seems to be rather unlikely. Nevertheless, we performed a search, similar to the one for the Fano plane and the Mermin square, within the cartographic $C_2^+$. Using Magma, we found in it $5916$ groups of index ten,  
$14$ of which have their coset groups isomorphic to $S_5$. However, we checked that maximum three (of five) lines of the pentagram are stabilized. This implies the non-existence of the Belyi map and puts three-qubit contextuality on a qualitatively different footing when compared with the two-qubit case.

Let us conclude this essay by an excerpt from Lewis Carroll's tale: \lq The hunting of the snark'

 \lq\lq {\it What's the good of Mercator's North Poles and

Equators,

Tropics, Zones, and Meridian Lines?"

So the Bellman would cry: and the crew would reply

\lq\lq They are merely conventional signs!}

\small 
\section*{Acknowledgements}
I thank L. Schneps and P. Lochak for inviting me to attend the Luminy school on \lq\lq The Grothendieck Theory of Dessins d'Enfants", held April 1993, and A. Giorgetti for his current collaboration concerning the enumeration of maps on surfaces. Some ideas presented above originated from lively discussions with my colleagues M. Saniga and P. L\'evay at the Mathematisches Forschungsinstitut Oberwolfach (Oberwolfach, Germany), within a {\it Reseach in Pairs} programme.


\end{document}